\documentclass[fleqn,usenatbib,onecolumn]{mnras}
\usepackage{newtxtext,newtxmath}
\usepackage[T1]{fontenc}
\DeclareRobustCommand{\VAN}[3]{#2}
\let\VANthebibliography\thebibliography
\def\thebibliography{\DeclareRobustCommand{\VAN}[3]{##3}\VANthebibliography}

\usepackage{amsmath}
\usepackage{graphicx}
\usepackage{url}
\usepackage{xcolor}
\usepackage{pdflscape}
\usepackage{geometry}      
\usepackage{color}

\newcommand{\beq}{\begin{equation}}
\newcommand{\eeq}{\end{equation}}
\newcommand{\beqa}{\begin{eqnarray}}
\newcommand{\eeqa}{\end{eqnarray}}

\def\la{\lower.5ex\hbox{$\; \buildrel < \over \sim \;$}}
\def\ga{\lower.5ex\hbox{$\; \buildrel > \over \sim \;$}}

\title{The Extended Plane of the Local Supercluster} 
\author[P. J. E. Peebles]{P. J. E. Peebles\\
Joseph Henry Laboratories, Princeton University, Princeton, NJ 08544, USA\\
{\rm pjep@Princeton.edu}}

\begin{document}
\label{firstpage}
\maketitle

\begin{abstract}\noindent
An update of the evidence that radio galaxies and clusters of galaxies are more common than average near the plane of the de Vaucouleurs Local Supercluster shows that in the distance range 100 to 200Mpc objects whose positions are correlated with the plane of the Local Supercluster include galaxies that are exceptionally luminous at two microns,  radio galaxies, and clusters of galaxies. There can be little doubt about this property of  cosmic structure. I also argue for detection of this correlation for the galaxies at 400Mpc distance that are exceptionally luminous at two microns. It will be interesting to learn whether these results are expected in the standard cosmology.
\end{abstract}

\section{Introduction}\label{sec:1}

It is broadly agreed that the dark sector of the standard $\Lambda$CDM cosmology seems likely to be too simple. If this or any other aspect of the theory differs from reality enough to matter it means there are anomalies to be found in the comparisons of observations with what the theory predicts. The observations considered here are the distances at which extragalactic objects are observed to be more common than average close to the extended plane of the Local Supercluster. The measure of the effect used in this paper is simple and predictive; it is intended to make the results convenient for comparisons to simulations of the growth of cosmic structure out of the initial conditions of the $\Lambda$CDM theory. 

Reviews of the literature on the Local Supercluster (or Virgo Supercluster, but hereinafter named the LSC) are presented by de Vaucouleurs (1958, 1978), Oort (1983), Tully (1986), Sanders and Mirabel (1996), McCall (2014) and others. De Vaucouleurs, de Vaucouleurs and Corwin (1976) gave the present standard direction of the supergalactic pole, 
\beq
l = 47.37^\circ, \  \ b = 6.32^\circ, \label{eq:LSC pole} 
\eeq
in galactic coordinates. The plane of the LSC is normal to this pole and passes through our Milky Way galaxy. The evidence that this direction is well chosen is the distributions of distances SGZ of objects from the plane of the LSC with distinctive peaks at SGZ close to zero. Examples are in de Vaucouleurs (1978) Figure 1; Tully (1982) Figure 4; and Zel’dovich, Einasto, and Shandarin (1982) Figure 5. Shaver's (1991) Figure 5 is an elegant way to illustrate the distribution of distances SGZ of galaxies from the plane of the LSC and the positions projected on the plane. Shaver used this figure to show that radio galaxies out to 85Mpc distance are distinctly closer to the plane of the LSC than are normal $L\sim L_\ast$ galaxies. These phenomena are worth closer study. To what distance is there evidence of the correlation of positions of extragalactic objects with the plane of the LSC? Shaver found that galaxies that are exceptionally luminous at radio wavelengths are particularly well correlated with the plane of the LSC. Are galaxies that are exceptionally luminous at the wavelengths typical of starlight particularly well correlated too? 

It important to know what the standard $\Lambda$CDM theory predicts. This has been addressed by, among others, Neuzil, Mansfield, and  Kravtsov (2020); Aragon-Calvo, Silk, and Neyrinck (2023); Schaller (2024);  and Wempe, White, Helmi, et al. (2026). The statistical measure presented here is meant to aid increased distances to which $\Lambda$CDM predictions can be obtained and compared to what is observed.  

In Peebles (2023) I argued that ``the measures in the bin $0.042 < z < 0.085$ [or $\sim180$ to 360 Mpc] provide a serious addition to the evidence of the presence of Sheet A as a continuation of the de Vaucouleurs Local Supercluster.'' The problem with this result is that I allowed the direction of maximum count in a postulated sheet-like distribution to bend with increasing distance. The postulated bends are modest, as illustrated by Figures 12 and 13 in Peebles (2023), but the intuitive judgement of what is a modest bend is not readily applied to the comparison of numerical simulations to observations. I am indebted to John Peacock and Kate Storey-Fisher for convincing me that determining the possible significance of evidence of extension of the plane of the Local Supercluster is best done using a simple statistical measure with minimal adjustable parameters that allow direct tests. The purpose of this paper is to present an example. 

The statistical measure for this study is discussed in Section~\ref{sec:the measure}. A summary of the provenance of the data used might be helpful; it is presented in Section~\ref{sec:data}.  Section~\ref{sec:results} shows results of the application of the statistical measure to the data and Section~\ref{sec:mocks} offers measures that might be useful for comparisons to what is found in mock catalogs. It is natural to wonder whether a sheet near which extragalactic objects are more common than usual is bent, close to the plane of the Local Supercluster nearby, as observed, but bent away from the LSC further out. A brief example in presented in Section~\ref{sec:bent}. Summary thoughts are in Section~\ref{sec:conclusions}.

\section{The Statistical Measure}\label{sec:the measure}

For the purpose of detecting a possible correlation of positions of extragalactic objects with the plane of the Local Supercluster, the LSC, and comparing the results to what is predicted by $\Lambda$CDM or some other theory, it is best to use a statistic with as few adjustable parameters as possible. This, with the evidence reviewed above that the direction of the LSC pole in equation~(\ref{eq:LSC pole}) is a good approximation, motivates taking this direction to be fixed by equation~(\ref{eq:LSC pole}). Then the judgement call is whether angular distributions of objects in disjoint distance bins show the appearance of significant correlation with this given orientation. The appearance is expected at relatively small distances where observations were used to fix the pole. The appearances become tests of predictions at greater distances when applied to catalogs that were not available when de Vaucouleurs et al. (1976) set the direction. 

The measure of angular distribution used here is the distribution of sin(SGB), the sine of the supergalactic latitude SGB of an object. (SGB is the angle from the LSC plane to the direction to the object relative to the Milky Way, positive if above the plane in the direction of the pole, negative if below. The angular distance $\theta$ of the object from the pole is $\theta = 90^\circ -$SGB. This is a return to the statistic I applied in Peebles 2022a and 2022b.) If the objects in a full sky sample are distributed as an isotropic random process the expected distribution of values of sin(SGB) is uniform between $-1$ and $+1$. A tendency of the objects to be close to the LSC plane would be observed as a maximum in the distribution of sin(SGB) at SGB close to zero. A tendency for galaxies to be close to a plane with nearest distance $h$ from the Milky Way would place these galaxies at $\hbox{sin(SGB)} \sim\pm 1$ if at galaxy distance close to $h$, and $\hbox{sin(SGB)}\sim\pm h/d$ if at distance $d$ much larger than $h$, where one would say the galaxies are close to the plane of the LSC. The value of sin(SGB) does not depend on the distance, which can be an advantage because an error in catalog distances is an error in the characteristic scale at which the correlation with the LSC is probed, but the evidence of the correlation is preserved. Thus this statical measure is useful for detection of whether there are more objects than average near the plane of the LSC that contains the Milky Way.  

Obscuration near the plane of the Milky Way galaxy usually prevents a full sky sample. This is taken into account by removing catalog members that are too close to the plane of the Milky Way, with a numerical computation of the mean effect on the distribution of sin(SGB) for an isotropic random process.

If objects in a sample are more numerous than average near the extended plane of the LSC it need not be statistically significant when considered alone. It is a significant contribution to the evidence of correlation if the appearance is present in other independent samples. The judgement of the number of appearances that makes a persuasive case for a real correlation can only be intuitive, but we are accustomed to intuitive assessments of this sort in science. I will argue that the appearances of maxima near zero in the measured distributions of sin(SGB) found from a range of distances and kinds of objects are abundant enough to make a persuasive case for detection of this measure of cosmic structure in the distance bin 100 to 200Mpc, with a good case for detection out to 400Mpc.

\section{The data}\label{sec:data}

The data used for this discussion are (1) the Karachentsev, Makarov, and Kaisina (2013) catalog of nearby galaxies as presented in the NASA Updated Nearby Galaxy Catalog NEARGALCAT; (2) the Huchra, Macri, Masters et al. (2012) galaxy redshift catalog based on the K-band 2MASS sky survey (Skrutskie, Cutri, Stiening, et al. 2006), excluding galaxies with redshift errors $cz$ not given or larger than 100 km~s$^{-1}$, hereinafter referred to as 2MRS galaxies; (3)  the Saunders, Sutherland, Maddox, et al. (2000) catalog of galaxies identified as point sources in the IRAS sky survey  (Beichman, Neugebauer, Habing, et al. 1985) at infrared wavelengths, hereinafter named PSCZ galaxies; (4) the van Velzen, Falcke, Schellart, et al. (2012) radio galaxy redshift catalog; (5) the Abell clusters of galaxies (Abell, Corwin, and Olowin 1989) updated and with measured redshifts in ABELLZCAT; and (6) the Klein, Hern\'andez-Lang, Mohr, et al. (2023, 2024) redshift catalog of clusters of galaxies detected by the X-ray emission by intracluster plasma, from the ROSAT satellite mission with a considerable variety of other programs to confirm detections and establish cluster redshifts.

I use Hubble constant $H=70\hbox{ km s}^{-1}~\hbox{ Mpc}^{-1}$ to convert redshifts to distances. At the greatest distances discussed, redshifts $z\sim 0.1$, the relativistic corrections to distances and absolute magnitudes are small and ignored here. 

I take account of the zone of avoidance by removing radio galaxies closer than $5^\circ$ from the galactic plane, following van Velzen et al. (2012). In other samples of galaxies and clusters of galaxies I remove objects at galactic latitudes lower than $20^\circ$. The effect on the distribution of sin(SGB) expected for an isotropic sample is computed by numerical random sampling. Because the plane of the LSC is close to perpendicular to the plane of our galaxy (the pole of the LSC is at galactic latitude $6.32^\circ$) the largest effect of the zone of avoidance is the suppression of counts near sin(SGB)$=\pm 1$. 

I show evidence that galaxies that are exceptionally luminous at wavelengths characteristic of starlight are more tightly correlated with the extended plane of the LSC than are $L\sim L_\ast$ galaxies. The standard for exceptionally luminous is the 50 most luminous 2MRS galaxies in a distance bin. The histograms showing distributions of sin(SGB) for less luminous galaxies are normalized by the factor that brings the count under the histogram to 50, which offers ready comparison to the distributions of sin(SGB) for the most luminous galaxies. 

The parameters that remain free after these considerations are the choices of bins of distance and luminosity. These limited choices simplify intuitive judgements of credibility of detection of alignment with the LSC and, I hope, simplify assessments to come from comparisons to reliable mock catalogs.

\begin{figure}
\includegraphics[width=6.in]{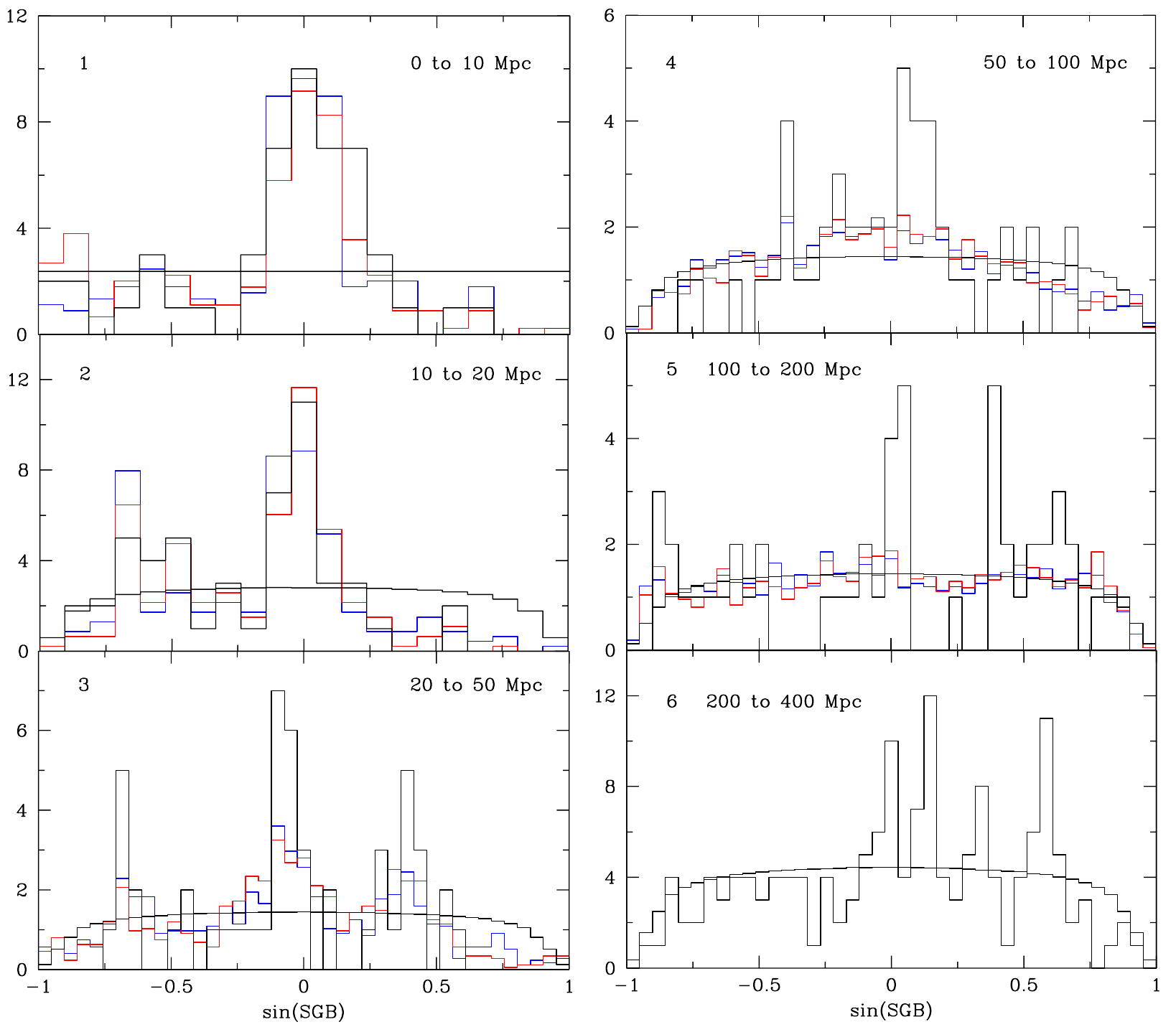}
\begin{centering}
\caption{Measures of the correlation of positions of 2MRS galaxies with the plane of the Local Supercluster. In all but panel 6 the black histogram with peaks is the distribution of sin(SGB) for the 50 most luminous  galaxies. The red histogram is the distribution of sin(SGB) for the galaxies with intermediate luminosities and the blue histogram the distribution for the remaining galaxies that are luminous enough for catalog completeness, both scaled to 50 under the histogram. The smooth histogram is the mean distribution for an isotropic process.} 
\label{fig:2MRSgalaxies}
\end{centering}
\end{figure}

\section{Results}\label{sec:results} 

The Huchra et al. (2012) catalog of 2MRS galaxies, with the Karachentsev et al. (2013) catalog of nearby galaxies, offers the particularly rich range of evidence of alignment with cosmic structure presented in Section~\ref{subsec:2MRS}. It is important that redshift catalogs of other extragalactic objects give evidence of consistency that add to the weight of evidence of the correlation. This is reviewed in Section~\ref{consistency checks}.

\subsection{The 2MRS redshift sample}\label{subsec:2MRS}

Figure~\ref{fig:2MRSgalaxies} shows frequency distributions of sin(SGB) for 2MRS galaxies in six disjoint distance bins and, apart from Panel~6, at each distance bin three disjoint bins in luminosity. The ranges of luminosities are chosen to aid the search for a possible relation between the galaxy luminosities and the degree of correlation of positions with the direction of the plane of the  LSC. This follows Shaver's (1991) demonstration that galaxies that are exceptionally luminous at radio wavelengths are exceptionally close to the plane. 

A 2MRS catalog completeness limit is K$_{\rm s}$-band apparent magnitude 11.75. Apart from Panel 1 an absolute magnitude for completeness in a distance bin is computed from this apparent magnitude at the greatest distance in the bin. In Panel~1 the absolute magnitude limit for completeness is taken to be $-15.0$. The Karachentsev et al. (2013) catalog used for Panel 1 contains many less luminous galaxies but the distances tend not to reach 10Mpc, an indication of incompleteness.

The black histogram with peaks in each panel but 6 is the distribution of sin(SGB) for the 50 most luminous galaxies in the distance bin. In Panel 6 the black histogram is the distribution of the 154 galaxies that are more luminous than the completeness limit.  The galaxies that are less luminous than the top 50 but more luminous than the completeness limit are placed in two groups of equal number (maybe apart from a single galaxy). The red histograms are the distributions of sin(SGB) for the more luminous half of these galaxies, and the blue histograms the distributions for the less luminous half. In all panels but 6 the red and blue histograms are scaled to a count of 50 each.

The Karachentsev et al. (2013) catalog of nearby galaxies is used in Panel 1 in  Figure~\ref{fig:2MRSgalaxies} for galaxies closer than 10Mpc. At this distance galaxies are detected at low galactic latitudes so the histograms in Panel~1 include all galactic latitudes. The other five panels use the Huchra et al. (2012) catalog of 2MRS galaxies. The effect of the zone of avoidance in these five distance bins is approximated by excluding objects closer than $|b| = 20^\circ$ to the galactic equator. The smooth histogram in each of these panels is the mean distribution of sin(SGB) for a statistically isotropic distribution with the appropriate  cut at low galactic latitude. 

The first panel uses catalog distances. In the other panels the galaxies are sorted in distance by the catalog redshifts with $H=70\hbox{ km s}^{-1}\hbox{ Mpc}^{-1}$, without relativistic correction. Karachentsev et al. (2013) give B-band absolute magnitudes of nearby galaxies, but for consistency I use absolute magnitudes computed from the catalog distances and the Huchra et al. (2012)  apparent magnitudes. 

In Panel 1 in Figure~\ref{fig:2MRSgalaxies}, for galaxies closer than 10Mpc, the black histogram is the distribution of sin(SGB) for the 50 galaxies that are more luminous than K-band absolute magnitude $-20.90$, the red histogram for the distribution of the 224 galaxies with absolute magnitudes between that and $-16.83$, and the blue histogram for the distribution of the 223 galaxies at $-16.83$ to $-15.00$. (Absolute magnitudes computed from the Huchra et al. 2012  apparent magnitudes are presented to two places after the decimal. They might not be this accurate but it aids checks if the reader is so inclined. In some cases even more decimal places are needed to get the wanted round numbers of counts, and they are best found by trial.) The straight line in Panel~1 is the mean for 50 objects isotropically distributed, to be compared to the distributions in the three luminosity bins.  

The distinct peaks at low SGB in the distributions of the three luminosity groups of nearby galaxies in Panel~1, distances less than 10Mpc, mean that significant numbers of the galaxies are closer to the plane of the LSC than they are distant from us. This is not new. A much earlier example is Figure~1 in de Vaucouleurs (1978 p. 205) and the statement there that the ``dwarf elliptical galaxies of the Sculptor type are also strongly concentrated toward the supergalactic equator especially towards the NGH [northern galactic hemisphere] (Karachentseva 1969), as are the DDO [David Dunlap Observatory] dwarfs, and more generally the low-velocity galaxies of all types." The similarities of the three histograms in Panel 1 agree with the observation that on large scales galaxies from giants to dwarfs have similar space distributions. Here we see a tendency to be aligned with the plane of the LSC in the same way for large and small galaxies.

The absolute magnitudes and counts of galaxies mentioned here for 2MRS galaxies closer than 10Mpc, and the same quantities for more distant samples, are listed in Table~\ref{tab:2MRScounts&luminosities}. The first absolute magnitude in each row is the lower bound of the luminosities of the 50 most luminous galaxies, the second the lower bound on the luminosities of the intermediate group, with the distribution of sin(SGB) plotted in red in Figure~\ref{fig:2MRSgalaxies}, and the third the lower bound on luminosities of the disjoint group set by the completeness limit and plotted as the blue histogram. The number of galaxies in each of the two groups less luminous than the top 50 is entered  in the fifth column. The sixth  column is the number density of the 50 most luminous galaxies in the bin, and the seventh the number density in each of the less luminous groups.

\begin{table}
\begin{centering}
\caption{2MRS galaxy counts and luminosities}
\medskip
\label{tab:2MRScounts&luminosities}
\begin{tabular}{cccccccc} 
\hline \hline
distances$^\ast$ & \multicolumn{3}{c}{magnitude bounds}& \multicolumn{1}{c}{count} & \multicolumn{2}{c}{number densities$^\ast$} \\
\hline
0 to 10 & $-20.90$ & $-16.83$ & $-15.00$ &  224  & $1.8\times 10^{-2}$ & $8.1\times 10^{-2}$  \\
10 to 20& $-23.76$ & $-21.63$ & $-19.76$ & 232  & $2.6\times 10^{-3}$ & $1.2\times 10^{-2}$ \\
20 to 50& $-24.95$ & $-22.98$ & $-21.74$ & 876 & $1.6\times 10^{-4}$ & $2.7\times 10^{-3}$ \\
50 to 100& $-25.59$ & $-23.89$ & $-23.25$ & 2903 & $2.1\times 10^{-5}$ & $1.2\times 10^{-3}$ \\
100 to 200& $-26.04$ & $-25.01$ & $-24.76$ & 2340 & $2.6\times 10^{-6}$ & $1.2\times 10^{-4}$ \\
200 to 400 &   ---   & ---  & $-26.26$ & 154 & $3.2\times 10^{-7}$ & $1.0\times 10^{-6}$ \\
\hline
\multicolumn{6}{l}{$^\ast$ the length unit is megaparsecs}\\
\end{tabular}
\end{centering}
\end{table}
 
In Panel~2 in Figure~\ref{fig:2MRSgalaxies}, distance bin 10 to 20Mpc, the distributions of sin(SGB) in the three luminosity classes are again quite similar.  A more modest peak in the distributions near sin(SGB)$\sim-0.5$ might be of some significance or might only be an accident of the clumpy  galaxy space distribution. The same could be true of the peaks at sin(SGB) close to zero, of course, but the coincidence of similar distributions in the three groups of luminosity in this distance bin with the peaks in the distributions of sin(SGB) for the galaxies closer than 10Mpc adds to the evidence that galaxies at this range of distances tend to be significantly closer to the plane of the LSC than expected from statistical isotropy.

At 20 to 50Mpc the larger numbers of galaxies allow smaller intervals of sin(SGB) in the histograms in Panel~3 and the rest of the panels. The distribution of sin(SGB) for the 50 most luminous galaxies in Panel 3 has three distinct peaks. As before, the two at sin(SGB)$\sim\pm0.5$ might signify something interesting, but I do not know anything to suggest. The appearance of the central peak of the black histogram where it is expected contributes to the evidence of the correlation of positions of particularly luminous galaxies with the direction set by the LSC. The similar smaller central peak counts of the less luminous galaxies in the red and blue histograms might signify a transition to the situation illustrated in the next two panels. 

In Panel~4, distances 50 to 100~Mpc, the histograms of 2903 galaxies each, red and blue, are similar and systematically larger than expected from isotropy across the central part of the distribution in sin(SGB). This might be a remnant of the tendency of galaxies large and small to align with the LSC. But it is notable that the central peak in the black histogram, for the 50 most luminous galaxies with luminosities greater  than $-25.59$, indicates that the positions of these exceptionally luminous galaxies are more likely to be close to the plane of the LSC than are the galaxies more luminous than absolute magnitude $-23.89$ plotted in the red histogram, and the blue histogram for the galaxies more luminous than the completeness bound $-23.25$.

The black histogram for the 50 most luminous galaxies in Panel 5, distances 100 to 200Mpc, has peak counts near sin(SGB) = 0.05, close to where it is expected, and  at sin(SGB) $\sim  0.39$. As before it cannot be determined whether either peak is significant from this panel alone, but the appearance of the peak at SGB near zero adds to the evidence of alignment with the plane of the LSC. The red and blue histograms are quite similar to each other and to the smooth histogram representing the mean for an isotropic random process. By this measure galaxy clustering averages out to isotropy on the scale of 200Mpc, apart from the correlation of positions with the LSC of the 50 exceptionally luminous galaxies at 100 to 200Mpc.

At distances 200 to 400Mpc, redshifts $z= 0.047$ to 0.093, 154 2MRS galaxies are luminous enough for completeness. The 50 most luminous  are brighter than absolute magnitude $-26.47$, and the completeness limit is $-26.26$. Since this is scarcely fainter the difference of  distributions of sin(SGB) for the top 50 and the next 104 is likely to be largely noise rather an effect of luminosity. The distribution of sin(SGB) for the 154 galaxies in Panel~6 shows peak counts at sin(SGB) = 0, 0.15, and 0.58. The first two have the appearance of a peak count where expected, making it an addition to the evidence that the orientation of the LSC correlates with cosmic structure beyond 200Mpc. The third peak must be considered noise unless some reason is found to think otherwise.

In the last line of Table~\ref{tab:2MRScounts&luminosities} the absolute magnitude is at the completeness limit at 400Mpc distance. The number densities are of the 50 most luminous and of the total of 154 galaxies more luminous than completeness. 

\begin{figure}
\begin{centering}
\includegraphics[width=6.in]{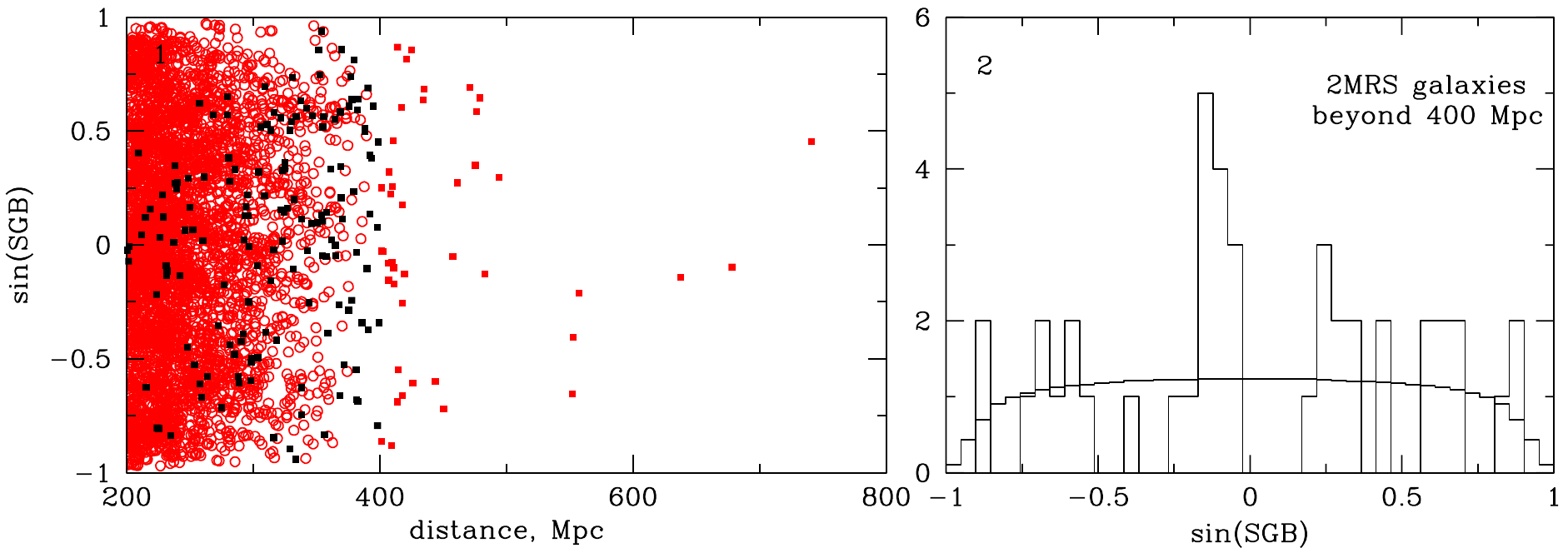}
 \caption{Distributions of distances and values of sin(SGB) of 2MRS galaxies at greater distances.}
 \label{fig:400to800_distribs}
 \end{centering}
\end{figure}

Evidence of correlation with the LSC of the 2MRS galaxies at distances of about 400Mpc is illustrated in Figure~\ref{fig:400to800_distribs}.  Panel 1 shows the joint distribution of galaxy distances and values of sin(SGB). The galaxies that are closer than 400Mpc and luminous enough for completeness are plotted as the filled black squares. The distribution of sin(SGB) for these galaxies, and the appearance of correlation with the orientation of the LSC, is shown in Panel~6 in Figure~\ref{fig:2MRSgalaxies}. The close to uniform distribution of distances of these black circles argues for a reasonably uniform sampling of sin(SGB) at distances 200 to  400Mpc. The open red circles show positions of the 2MRS galaxies that are less luminous than the completeness limit. They illustrate the decreasing numbers of detections of less luminous galaxies with increasing distance, and that all 43 2MRS galaxies with catalog distances greater than 400Mpc are luminous enough for completeness at their catalog distances. The most distant of the 2MRS galaxies are at 637, 678, and 741Mpc. The majority are close to 400Mpc where catalog distance errors could have placed some on the wrong side of 400Mpc. 

Panel 2 in Figure~\ref{fig:400to800_distribs} shows the distribution of sin(SGB) for the 43 galaxies at catalog distances greater than but  largely close to 400Mpc. The appearance of a peak count close to sin(SGB) = 0 adds to the evidence that the correlation of positions of galaxies with the plane of the LSC is detected to about 400Mpc in the 2MRS data, but not much farther than that.

\begin{figure}
\begin{centering}
\includegraphics[width=6.in]{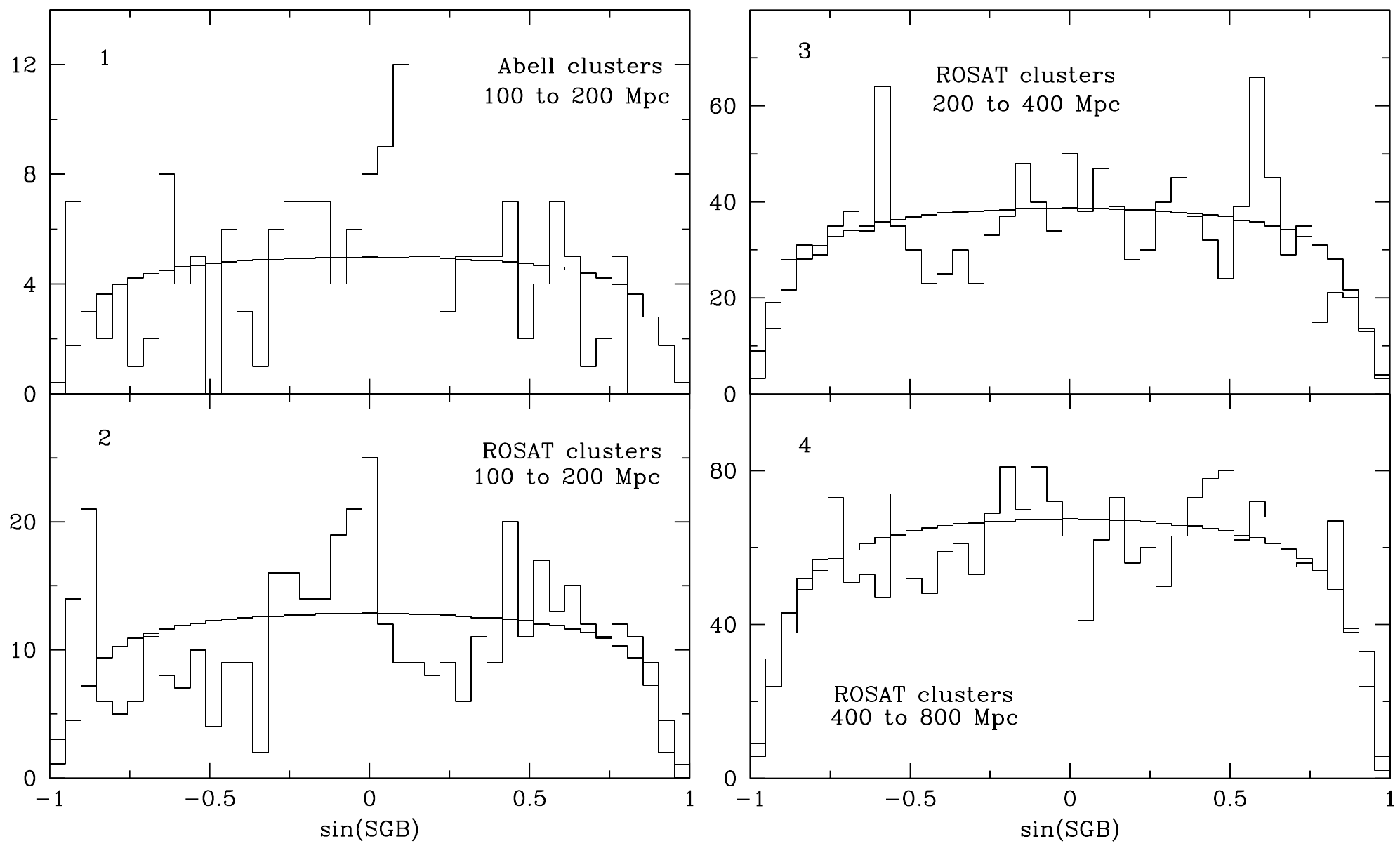}
 \caption{Distributions of sin(SGB) for clusters of galaxies.}
 \label{fig:clusters}
\end{centering}
 \end{figure}

\subsection{Checks from distributions of radio galaxies and clusters of galaxies}
 \label{consistency checks}

Many of the more prominent nearby clusters of galaxies are close to the plane of the LSC; familiar examples are the Virgo cluster at about 17Mpc and supergalactic latitude $\sim -2^\circ$ and the Coma cluster at about 110Mpc and supergalactic latitude $\sim 8^\circ$. The situation for other more distant clusters is illustrated in Figure~\ref{fig:clusters}. The clusters in this sample are those farther than $20^\circ$ from the plane of the Milky Way. 

Panels 1 and 2 in Figure~\ref{fig:clusters}  show the distributions of sin(SGB) at distances 100 to 200Mpc for the 173 Abell clusters in the ABELLZCAT catalog and the 446 ROSAT clusters in the Klein et al. (2024) catalog (with number densities $9.0\times 10^{-6}$ and $2.3\times 10^{-5}$Mpc$^{-3}$). These clusters were identified in quite different ways, Abell clusters by visual inspections of photographic plates, ROSAT clusters by satellite detections of X-rays from hot intracluster plasma. Both are catalogs of clusters of galaxies, but a catalog inevitably has effects of systematic errors, and it is conceivable that some systematic error causes an artificial excess of counts at low SGB in one of the catalogs. But the effects of systematic errors in one sample are not likely to resemble the effects in the other sample that was obtained in such a different way. Thus it is an important check that the clusters identified these two different ways have similar distributions of sin(SGB) with similar appearances of a peak count near SGB = 0. And the agreement with the near central peak in the distribution of 2MRS galaxies in Panel~5 in Figure~\ref{fig:2MRSgalaxies} is an important addition to the evidence that cosmic structure is correlated with the LSC at 100 to 200Mpc.

The modest excess of counts over the mean near SGB = 0 in Panel 3 in Figure~\ref{fig:clusters}, for 1342 ROSAT clusters in the distance bin 200 to 400Mpc, seems too weak for further consideration. More striking is the appearance of near symmetry of reflection through the plane of the LSC here and in Panel~3 in Figure~\ref{fig:2MRSgalaxies}. I suppose both are accidental. More directly relevant is that the 2343 ROSAT clusters at 400 to 800Mpc do not indicate correlation of cluster positions with an extended flat plane of the LSC.

 \begin{figure}
\begin{centering}
\includegraphics[width=6.in]{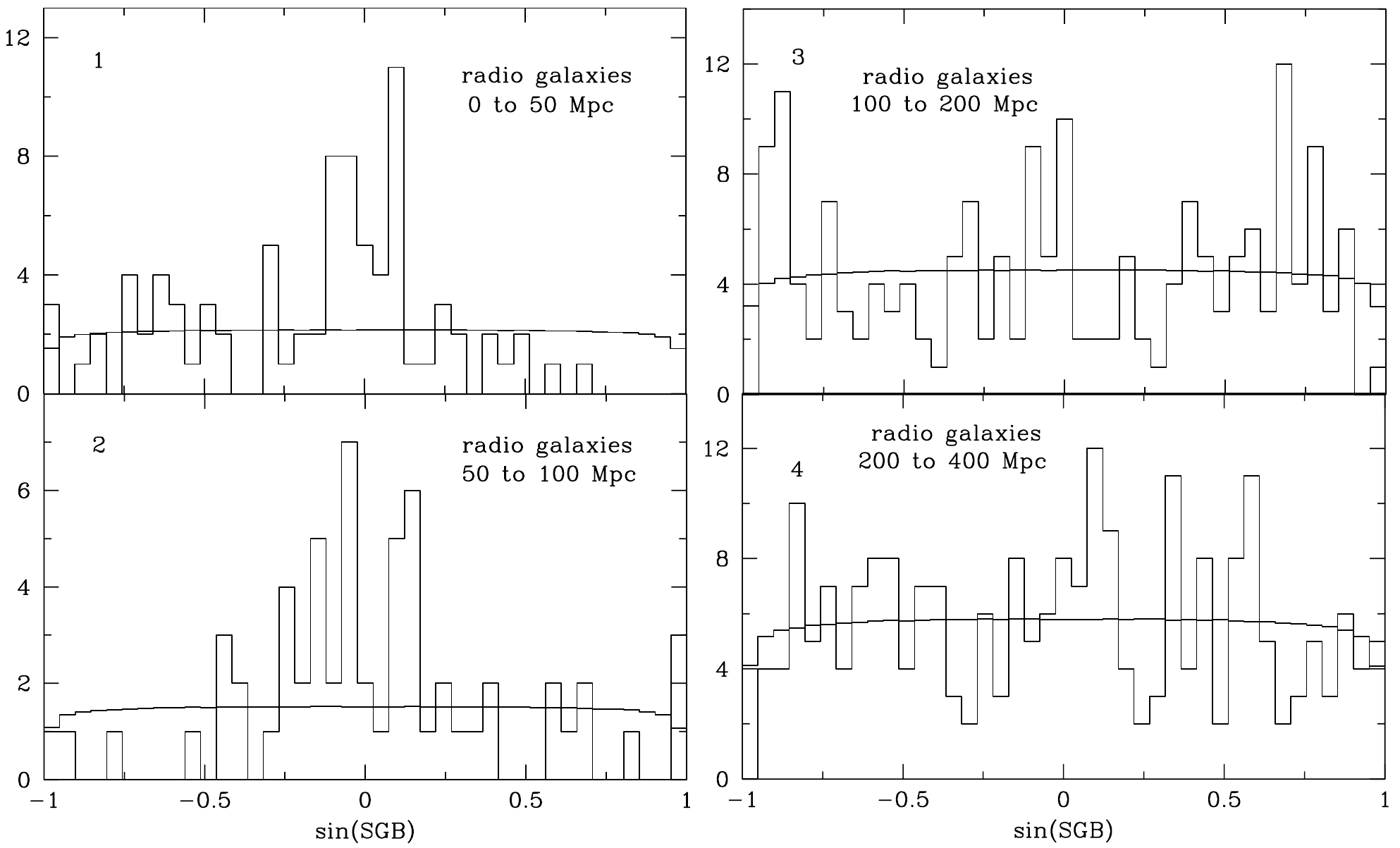}
 \caption{As in Figure~\ref{fig:clusters} for radio galaxies.}
 \label{fig:radio}
\end{centering}
 \end{figure}
 
Shaver (1991) demonstrated the correlation of positions of radio galaxies with the plane of the LSC out to about 85Mpc and, equally interesting, the lack of this correlation for $L\sim L_\ast$ galaxies. Figure~\ref{fig:radio} shows the situation for the van Velzen et al. (2012) radio galaxies. No luminosity cut is applied to this sample and, following van Velzen et al., sources closer than $5^\circ$ from the galactic equator are rejected. Shaver's effect is clear for the 84 radio galaxies in panel~1 and the 58 in panel~2, at distances where Shaver had already demonstrated the correlation. Peak counts in the distributions of sin(SGB) near SGB = 0 also appear in Panel 3, for 165 radio galaxies at 100 to 200Mpc, and Panel 4, for 221 radio galaxies at 200 to 400Mpc. These peaks would not be of much interest if considered alone, but their  appearances where expected in all four panels in Figure~\ref{fig:radio} make a serious case for detection of the correlation of radio galaxy positions with the LSC beyond 200Mpc.

 \begin{figure}
\begin{centering}
\includegraphics[width=6.in]{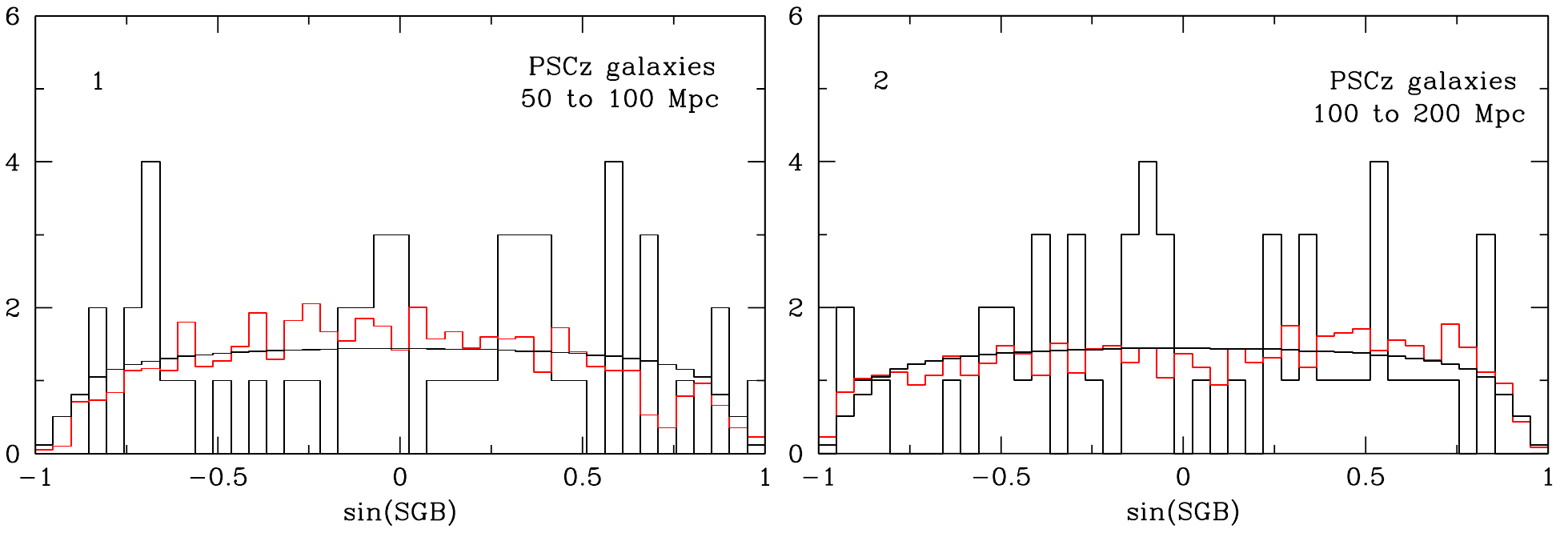}
 \caption{As in Figure~\ref{fig:clusters} for PSCZ galaxies.}
 \label{fig:PSCZ}
\end{centering}
 \end{figure}

The PSCZ catalog of galaxies discovered as point-like sources in the Infrared Astronomical Satellite survey of the sky at wavelengths ranging from 8 to $100\mu$, with redshifts cataloged by Saunders et al. (2000), quite uniformly covers almost the entire sky. The distributions of sin(SGB) in Figure~\ref{fig:PSCZ}  exclude galaxies below galactic latitude $20^\circ$. A  measure of the infrared luminosity of a PSCZ galaxy is the product $J = S_{60}D_{\rm kpc}^2$, where $S_{60}$ is the flux density in Jansky at $60\mu$ and $D_{\rm kpc}$ is the distance to the galaxy in kiloparsecs.  The black histograms in Figure~\ref{fig:PSCZ} are the distributions of sin(SGB) for the 50 most powerful sources of radiation at $\sim 60\mu$, with $J>0.042$Jy~kpc$^2$ in Panel~1 and $J>0.11$Jy~kpc$^2$ in Panel~2. The red histograms are the distributions of sin(SGB) for the galaxies with infrared luminosities between these bounds and a tenth of the bounds. There are 1972 of these galaxies in Panel~1 and 3083 in Panel~2. 

The similarity of the red histograms to the mean for an isotropic process illustrates the uniformity of the PSCz sky coverage and the approach to isotropy of the universe by this measure on the scale of 50 to 100Mpc. The probe for correlation at the slightly different direction in equation (11) in Peebles (2023) yields a more pronounced peak count in the black histogram in Figure~6 in this paper. One might argue for a modest case for the appearances of peak counts in the black histograms near SGB = 0 in both panels in Figure~\ref{fig:PSCZ}, but they are far less distinctive than for the 2MRS galaxies in Figure~\ref{fig:2MRSgalaxies}. A possible explanation is that PSCZ galaxies, which are selected because they are luminous in the infrared, are results of considerable bursts of star formation in ordinary large galaxies. The many young stars would produce abundant starlight and dust, and the dust would absorb the starlight and reradiate the energy in the infrared. In this picture the distribution of sin(SGB) for the top 50 PSCZ galaxies would not be expected to be  at strongly correlated with the LSC as the most luminous 2MRS galaxies. 

 \section{Data for comparisons to mock catalogs}\label{sec:mocks}

The results in Figure~\ref{fig:2MRSgalaxies} with the tests of consistency in Figures~\ref{fig:400to800_distribs} to \ref{fig:radio} can be used to test the $\Lambda$CDM theory: do distributions of objects in mock catalogs predicted by this theory contain similar examples?  For this purpose a  measure of central counts in excess of random in distributions of sin(SGB) can be defined as follows. 

Many of the histograms of sin(SGB) in Section~\ref{sec:results} have a central bin and 20 bins on either side,  41 at $j = -20$ to $+20$. A measure of the central peak in the distribution of sin(SGB) is the sum $S$ of the counts in the central seven bins at $j = -3$ to $j =+3$, though trials might indicate a narrower or broader window. The mean sum in this window expected from  isotropy is $R$. An indication of the tendency of positions of objects to be closer than average to the plane of the LSC is the central contrast
\beq
C = {S\over R} - 1.\label{eq:contrast}
\eeq

\begin{table}
\begin{centering}
\caption{Measures for comparisons to mock catalogs}
\medskip
\label{tab:formocks}
\begin{tabular}{cccc} 
\hline \hline
 \multicolumn{1}{c}{distances$^\ast$} & \multicolumn{1}{c}{number density$^\ast$} & \multicolumn{2}{c}{central contrasts $C$}\\
 && \multicolumn{1}{c}{top 50} &  \multicolumn{1}{c}{next 100}\\
\hline
0 to 10  & $1.8\times 10^{-2}$   & 3.33  & 3.04\\
10 to 20 &  $2.6\times 10^{-3}$ & 1.28  & 1.48  \\
20 to 50  &  $1.6\times 10^{-4}$& 0.89 & 0.84 \\
50 to 100 & $2.1\times 10^{-5}$& 1.09 & 0.59 \\
100 to 200 & $2.6\times 10^{-6}$&  0.29 & $-0.20$\\
200 to 400 & $3.2\times 10^{-7}$ & 0.49 & 0.49 \\
\hline
\multicolumn{4}{l}{$^\ast$ the length unit is megaparsecs}\\
\end{tabular}
\end{centering}
\end{table}

Table~\ref{tab:formocks} shows quantities that might be useful for analyses of mock catalogs. They are derived from the Huchra et al. (2012) catalog with the results illustrated in Figure~\ref{fig:2MRSgalaxies}. For convenience the number densities of the 50 most luminous galaxies in each distance bin are repeated from Table~\ref{tab:2MRScounts&luminosities}. Central density contrasts $C$ defined in equation~(\ref{eq:contrast}) are listed in the last two columns. The first is the contrast for the most luminous 50 galaxies, the second the contrast for the next 100 most luminous. One of the twelve central contrasts is negative, meaning fewer counts than expected from isotropy. This is in the distance bin 100 to 200Mpc, where the central contrast of the 50 most luminous 2MRS galaxies is positive but the smallest of the eleven positive contrasts. The value $C=0.29$ can be taken to be significant evidence of correlation with the LSC, because the appearance of correlation in this distance bin is found also in both of the catalogs of clusters and in the radio galaxy catalog. 

\section{A bent extension of the Local Supercluster?}
\label{sec:bent}

It is impressive that the angular position of the supergalactic pole in equation~(\ref{eq:LSC pole}) that de Vaucouleurs et al. (1976) presented a half century ago still is a successful predictor of the distributions of extragalactic objects out to 200Mpc and likely to 400Mpc. But it is easy to imagine modest bends of the plane to which cosmic structure is maximally correlated at these and greater distances. This thought changes the reasonably specific prediction tested here to a suggestion that is best used for  comparisons to distributions of model galaxies or clusters of galaxies in mock catalogs, which we do not yet have. Thus I confine these remarks to a single example of the effect of adjustment of the direction of the supergalactic pole.

\begin{figure}
\begin{centering}
\includegraphics[width=6.in]{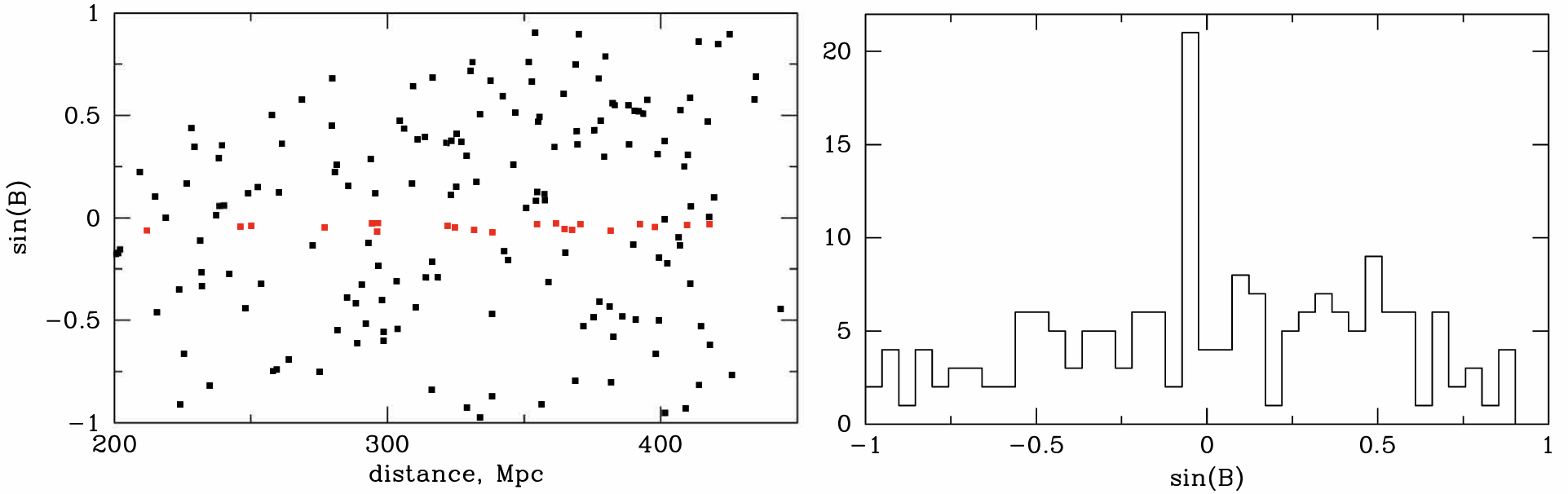}
\caption{An example of a flat sheet in the distribution of the 182 most luminous 2MRS galaxies at 200 to 450Mpc that is tilted $13^\circ$ from the pole of the LSC. The red points in the scatter plot are 
the 21 2MRS galaxies in the peak in the distribution of sin(B).}
 \label{fig:bend}
\end{centering}
 \end{figure}

This example uses the 2MRS galaxies at redshift distances 200 to 450Mpc, the upper limit allowing inclusion of the galaxies that in Figure~\ref{fig:400to800_distribs} are seen to be somewhat more distant than 400Mpc. The absolute magnitude cutoff, $-26.26$, is the apparent magnitude for completeness at 400Mpc. After exclusion  of the galaxies at galactic latitudes lower than $20^\circ$ the sample is 182 galaxies. Each iteration in the search for a flat sheet  of galaxies that passes near us begins with the orientation of a pole chosen isotropically at random. The pole defines a plane that passes through our position. The angular distances B of galaxies from the plane, meaning the latitudes relative to this pole, are sorted into counts of values of sin(B) in a histogram of 41 bins. The largest count among the 21 bins at $j = -10$ to +10 is found. This is restricted to the central region because I seek a modestly bent sheet. When a count in one of these 21 bins is larger than or equal to any previously found the count and direction of the pole are recorded. This is repeated $10^6$ times. The result always is a count of 21 of the 182 galaxies. The several ways to reach this count by slight changes of the direction of the pole place the peak count either at $j = -1$, as in the histogram in Figure~\ref{fig:bend}, or $j = +1$. The directions of the pole that get this largest count, 21, in the all-sky search, are in the range
\beq
l = 60.2\pm 0.2,\quad b = 9.8\pm 0.3 \quad \hbox{degrees}. \label{eq:bentdirection}
\eeq
This pole is tilted $13^\circ$ from the supergalactic pole in equation~(\ref{eq:LSC pole}). 

The scatter plot in Figure~(\ref{fig:bend}) shows the distances and values of sin(B) of the 182 galaxies with sin(B) computed from the direction in equation~(\ref{eq:bentdirection}). The 21 galaxies that are in the bin of sin(B) with the maximum count (here $j = -1$) are marked in red, the rest in black. Is the flat sheet of red points in Panel 1, along with the $22^{\rm nd}$, the Milky Way, a meaningless accident or maybe physically significant? Evidence for the latter is that the sheet is tilted from the plane of the Local Supercluster by only 13 degrees. Comparisons to mock catalogs without artificial regularities set by boundary conditions might yield a more definte answer. 

\section{Concluding remarks}\label{sec:conclusions}

The evidence of correlation of cosmic structure with the orientation of the Local Supercluster at 100 to 200Mpc distance appears in Panel~5 in Figure~\ref{fig:2MRSgalaxies} for 2MRS galaxies; in Panels 1 and 2 in Figure~\ref{fig:clusters} for the Abell and ROSAT clusters of galaxies that were identified by quite different methods; and in Panel 3 in Figure~\ref{fig:radio} for radio galaxies. If considered alone the evidence of a peak count near SGB = 0  in the distribution of sin(SGB) in a single panel need not be significant in many cases, particularly where there are other peaks larger than the central one. The critical point is the consistency of indications of peak counts at SGB close to zero from independently obtained tracers of cosmic structure. The amount of evidence of this kind required for a convincing conclusion must be a judgement call. Mine is that cosmic structure almost certainly is correlated with the plane of the Local Supercluster at distances of 100 to 200Mpc. 

The signature of correlation with the LSC at 200 to 400Mpc appears in this study only in the space distribution of 2MRS galaxies, as shown in Panel 6 in Figure~\ref{fig:2MRSgalaxies} and the two panels in Figure~\ref{fig:400to800_distribs}. They make a good case for detection at 400Mpc but not much beyond that. This result is not as well checked as at 100 to 200Mpc, but it is serious. 

At distances less than about 20Mpc the three distributions of sin(SGB) in the three bins of galaxy luminosity in each of Panels 1 and 2 in Figure~\ref{fig:2MRSgalaxies} look quite similar. By this measure positions of galaxies large and small are similarly correlated with the LSC. At distances greater than 50Mpc the larger space volumes allow larger samples of exceptionally luminous galaxies. The evidence in Panels 4 and 5 is that 2MRS galaxies  more luminous than about absolute magnitude $-26.0$ (in the Huchra et al. 2012 magnitude system) are special, more strongly correlated with the plane of the LSC than are less luminous galaxies. Panel 3, for distances 20 to 50Mpc, has the appearance of a transition from the closer  smaller samples with few exceptionally luminous galaxies to the situation in Panels 4 and 5. The lower bound on luminosity of the 50 most luminous galaxies is $-24.95$ in Panel~3. Perhaps this distance bin is close enough that galaxies in the red and blue groups are correlated with the LSC, but far enough that there are enough exceptionally luminous galaxies to produce a more prominent peak count at SGB close to zero in the distribution of sin(SGB) for the top 50 galaxies. 

The exceptionally luminous galaxies that are particularly well correlated with the plane of the LSC are mostly ellipticals (as indicated in Fig.~3 in Peebles 2022b). It would be interesting to see whether deeper images of these special galaxies reveal any other special properties. For example, the passages of close to straight cosmic strings moving at high speed through the early universe would produce sheets of matter that might produce sheets of galaxies, and these special galaxies might show signatures of the special way they formed. 

The PSCZ galaxies that are exceptionally luminous in the infrared are not as strongly correlated with the LSC as are the most luminous 2MRS galaxies selected at wavelengths typical of starlight. The easy explanation is that PSCZ galaxies typically have closer to normal stellar masses and are made luminous in the infrared by sporadic starbursts. But perhaps there is more to say about this.

It takes nothing from the skill and effort devoted to the compilation of the important 2MRS catalog to speculate that, among the 43533 catalog redshifts used here, a few are seriously wrong, the results of rare glitches in measurements or communications. Might the three most distant galaxies in the left panel in Figure~\ref{fig:400to800_distribs} have erroneous catalog redshifts? Deeper observations of these three galaxies might reveal that the distances are much smaller than the catalog values and these galaxies are not so special. But if the catalog distances of these three and the other galaxies that are indicated to be much farther than 400Mpc prove to be about right what might deeper images say about the properties of these exceptionally luminous galaxies? 

The purpose of this paper is to present data that can be compared to what is found in mock catalogs of the space distributions of model galaxies, or dark matter halos,  computed from the initial conditions of the standard $\Lambda$CDM theory. We do not have a predictive theory of why some dark matter halos contain galaxies that are extraordinarily luminous at radio wavelengths, so their distributions in Figure~\ref{fig:radio} serve mainly as additions to the evidence of correlation of cosmic structure with the orientation of the Local Supercluster. The evidence in Panels 4 and 5 in Figure~\ref{fig:2MRSgalaxies} is that galaxies with number densities on the order of $10^{-6}$Mpc$^{-3}$ are better correlated with the plane of the LSC than are  normal $L\sim L_\ast$ galaxies. Clusters of galaxies also are  better correlated with the plane, but their number density is an order of magnitude larger. Mock catalogs might help straighten this out and determine how frequently cosmic structure at 200Mpc aligns with structure analogous to the Local Supercluster on the scale of 20Mpc around analogs of the Milky Way. 

\section{acknowledgments} 
I am grateful to Kate Storey-Fisher and John Peacock for discussions that convinced me of the importance of mock catalogs to compare to what is found in real catalogs, to Renyue Cen for discussions of the art of constructing mock catalogs, and to Princeton University for its continued hospitality.

\end{document}